\documentstyle[aps,prd,eqsecnum,twoside]{revtex}
\newcommand {\ve}{\varepsilon}
\newcommand {\pr}{\partial}
\newcommand {\cG}{\cal G}
\newcommand {\cD}{\cal D}
\newcommand {\bg}{\bar \gamma}
\newcommand {\bp}{\bar \psi}

\begin{document}
\title{Dirac Spinor in Bianchi-I Universe with
time dependent Gravitational and Cosmological Constants}
\author{Bijan Saha\\ 
Laboratory of Information Technologies\\ 
Joint Institute for Nuclear Research, Dubna\\ 
141980 Dubna, Moscow region, Russia\\ 
e-mail:  saha@thsun1.jinr.ru, bijan@cv.jinr.ru}
\maketitle 
\thispagestyle{empty}
$ $
\vskip .1cm
\noindent
Self-consistent system of nonlinear spinor field  and Bianchi I (BI)
gravitational one with time dependent gravitational constant ($G$) and
cosmological constant ($\Lambda$) has been studied. The initial and the 
asymptotic behaviors of the field functions and the metric one have been 
thoroughly investigated. Given $\Lambda = \Lambda_0/\tau^2$, with 
$\tau = \sqrt{-g}$, $G$ has been estimated as a function of $\tau$.
The role of perfect fluid at the initial state of expansion and
asymptotical isotropization process of the initailly anisotropic
universe has been elucidated.  
\vskip 5mm

\section{Introduction}

Einstein's theory of gravity contains two parameters, considered as
fundamental constants: Newton's gravitational constant $G$ and the
cosmological constant $\Lambda$~\cite{einstein}. A possible time
variation of $G$ has been suggested by Dirac~\cite{dirac} and 
extensively discussed in literature~\cite{dicke,weinberg,wu,maeda,damour}. 
The ``cosmological constant'' $\Lambda$ as a function of time was  
studied by many authors. Chen and Wu~\cite{chen} advocated the
possibility that the cosmological constant varies in time as $1/R^2$, 
with $R$ being the scale factor of Robertson-Walker model. Further
Abdel-Rahman~\cite{rahman} considered a model with the same kind of 
variation, while Berman {\it et al.}~\cite{bermana,bermanb,som} stressed 
that the relation $R \propto t^{-2}$ plays an important role in
cosmology. Berman and Gomide~\cite{gomide} also showed that
all the phases of the universe, i.e., radiation, inflation, and 
pressure-free, may be considered as particular cases of the decelaration
parameter $q=$ constant type, where
\begin{equation}
q = -R {\ddot R}/{\dot R}^2,
\label{}
\end{equation}   
where dots stand for time derivative. This definition was extended by
Singh and Agrawal~\cite{singh} to the Bianchi cosmological models.
Recently we studied the behavior of self-consistent nonlinear spinor
field (NLSF) in a Bianchi I (B-I) universe~\cite{rybakov} that was 
followed by the study of the self-consistent system of interacting 
spinor and scalar fields~\cite{alvarado}. These studies were further
extended to more general NLSF in presence of perfect fluid
~\cite{saha1,saha2}.

The aim of this paper is to extend our study with time dependent
gravitational constant $G$ and cosmological constant $\Lambda$ in
Einstein's equation.
\vskip 5mm
\section{Fundamental equations and general solutions}
\setcounter{equation}{0}

The Dirac spinor field is given by the Lagrangian 
\begin{equation} 
L= \frac{i}{2} \biggl[ \bp \gamma^{\mu} \nabla_{\mu} \psi - 
\nabla_{\mu} \bp \gamma^{\mu} \psi \biggr] - m\bp \psi + L_N.
\label{splag}
\end{equation} 
The nonlinear term $L_N$ describes the self-interaction of 
a spinor field and can be presented as some arbitrary functions of 
invariants generated from the real bilinear forms of a spinor 
field having the form 
$$S\,=\, \bp \psi, \quad                    
P\,=\,i \bp \gamma^5 \psi, \quad
v^\mu\,=\,(\bp \gamma^\mu \psi), \quad
A^\mu\,=\,(\bp \gamma^5 \gamma^\mu \psi), \quad
T^{\mu\nu}\,=\,(\bp \sigma^{\mu\nu} \psi),$$
where $\sigma^{\mu\nu}\,=\,(i/2)[\gamma^\mu\gamma^\nu\,-\,
\gamma^\nu\gamma^\mu]$. Invariants, corresponding to the bilnear forms, 
look 
$$ I = S^2, \quad J = P^2, \quad 
I_v = v_\mu\,v^\mu\,=\,(\bp \gamma^\mu \psi)\,g_{\mu\nu}
(\bp \gamma^\nu \psi),$$ 
$$I_A = A_\mu\,A^\mu\,=\,(\bp \gamma^5 \gamma^\mu \psi)\,g_{\mu\nu}
(\bp \gamma^5 \gamma^\nu \psi), \quad
I_T = T_{\mu\nu}\,T^{\mu\nu}\,=\,(\bp \sigma^{\mu\nu} \psi)\,
g_{\mu\alpha}g_{\nu\beta}(\bp \sigma^{\alpha\beta} \psi).$$ 
According to the Pauli-Fierz theorem~\cite{landau}, among the five invariants
only $I$ and $J$ are independent as all other can be expressed by them:
$I_v = - I_A = I + J$ and $I_T = I - J.$ Therefore we choose the nonlinear
term $L_N = \lambda F(I, J)$, thus claiming that it describes the 
nonlinearity in the most general of its form. Here $\lambda$ is the coupling
constant.

The NLSF equations and components of the energy-momentum tensor for the 
spinor field corresponding to the Lagrangian (\ref{splag}) are
\begin{mathletters}
\label{speq}
\begin{eqnarray}
i\gamma^\mu \nabla_\mu \psi - \Phi \psi + i {\cG} \gamma^5 \psi\,&=&\,0, 
\label{speq1} \\
 i \nabla_\mu \bp \gamma^\mu +  \Phi \bp 
- i {\cG}  \bp \gamma^5 \,&=&\,0, \label{speq2}
\end{eqnarray}
\end{mathletters}
and
\begin{equation}
T_{\mu}^{\rho}=\frac{i}{4} g^{\rho\nu} \biggl(\bp \gamma_\mu 
\nabla_\nu \psi + \bp \gamma_\nu \nabla_\mu \psi - \nabla_\mu \bar 
\psi \gamma_\nu \psi - \nabla_\nu \bp \gamma_\mu \psi \biggr) \,-
\delta_{\mu}^{\rho}L_{sp}+ T_{\mu\,(m)}^{\rho},
\label{tem}
\end{equation}
where 
$$ \Phi = m - {\cD} = m - 2 \lambda S \frac{{\pr F}}{{\pr I}}, \qquad
{\cG} = 2 \lambda P \frac{{\pr F}}{{\pr J}}.$$

Here $T_{\mu\,(m)}^{\rho}$ is the energy-momentum tensor of a perfect fluid. 
For a Universe filled with perfect fluid, in the concomitant system of 
reference $(u^0=1, \, u^i=0, i=1,2,3)$ we have
\begin{equation}
T_{\mu (m)}^{\nu}\,=\, (p + \ve) u_\mu u^\nu - 
\delta_{\mu}^{\nu} p \,=\,(\ve,\,- p,\,- p,\,- p),
\end{equation} 
where energy $\ve$ is related to the pressure $p$ by the equation 
of state $p\,=\,\zeta\,\ve$. The general solution has been derived 
by Jacobs~\cite{jacobs}. Here $\zeta$ varies between the
interval $0\,\le\, \zeta\,\le\,1$, whereas $\zeta\,=\,0$ describes
the dust Universe, $\zeta\,=\,\frac{1}{3}$ presents radiation Universe,
$\frac{1}{3}\,<\,\zeta\,<\,1$ ascribes hard Universe and $\zeta\,=\,1$
corresponds to the stiff matter. In (\ref{speq}) and (\ref{tem}) 
$\nabla_\mu$ denotes the covariant derivative of spinor, having the 
form~\cite{zhelnorovich}  
\begin{equation} 
\nabla_\mu \psi=\frac{\pr \psi}{\pr x^\mu} -\Gamma_\mu \psi, 
\label{nab}
\end{equation} 
where $\Gamma_\mu(x)$ are spinor affine connection matrices.  

Einstein's field equations with variable cosmological and gravitational
``constants'' $\Lambda$ and $G$ are given by
\begin{equation}
R_{\nu}^{\mu} - \frac{1}{2}\,\delta_{\nu}^{\mu} R = - 8 \pi G(t) 
T_{\nu}^{\mu} + \Lambda(t) \delta_{\nu}^{\mu} 
\label{ee}
\end{equation} 
where $R_{\nu}^{\mu}$ is the Ricci tensor; $R = g^{\mu\,\nu} R_{\mu\,\nu}$
is the Ricci scalar; and $T_{\nu}^{\mu}$ is the energy-momentum tensor
of matter field given by (\ref{tem}). From the divergence of (\ref{ee}) 
we get
\begin{equation}
8 \pi G_{,\mu} T_{\nu}^{\mu} + 8 \pi G \bigl(T_{\nu;\mu}^{\mu}\bigr)
- \Lambda{,\mu} \delta_{\nu}^{\mu} = 0,
\label{div}
\end{equation} 

The Bianchi I model is given by
\begin{equation} 
ds^2 = dt^2 - a^{2}(t)\,dx^{2} - b^{2}(t)\,dy^{2} - c^{2}(t)\,dz^2.
\label{BI}
\end{equation}
We study the space-independent spinor fields, hence $T_{\nu}^{\mu}$
is the function of $t$ alone. Taking this into account 
for the metric (\ref{BI}), the Einstein's equations (\ref{ee}) and 
(\ref{div}) reduces to
\begin{mathletters}
\label{BID}
\begin{eqnarray}
\frac{\ddot b}{b} +\frac{\ddot c}{c} + \frac{\dot b}{b}\frac{\dot 
c}{c}&=&  8 \pi G T_{1}^{1} -\Lambda,\label{11}\\
\frac{\ddot c}{c} +\frac{\ddot a}{a} + \frac{\dot c}{c}\frac{\dot 
a}{a}&=&  8 \pi G T_{2}^{2} - \Lambda,\label{22}\\
\frac{\ddot a}{a} +\frac{\ddot b}{b} + \frac{\dot a}{a}\frac{\dot 
b}{b}&=&  8 \pi G T_{3}^{3} - \Lambda,\label{33}\\
\frac{\dot a}{a}\frac{\dot b}{b} +\frac{\dot b}{b}\frac{\dot c}{c} 
+\frac{\dot c}{c}\frac{\dot a}{a}&=&  8 \pi G T_{0}^{0} - \Lambda,
\label{00}
\end{eqnarray}
\end{mathletters}
\begin{equation}
8 \pi {\dot G} T_{0}^{0} + 8 \pi G \Bigl[{\dot T_{0}^{0}}
+ T_{0}^{0} \Bigl(\frac{\dot a}{a} + \frac{\dot b}{b} +
\frac{\dot c}{c}\Bigr) + T_{1}^{1}\frac{\dot a}{a}
+T_{2}^{2}\frac{\dot b}{b} + T_{3}^{3}\frac{\dot c}{c}\Bigr] - 
{\dot \Lambda} = 0, 
\label{divI}
\end{equation}
where points denote differentiation with respect to t. 
If we suppose the energy conservation law $T_{\nu;\mu}^{\mu} = 0$ to hold,
then (\ref{divI}) reduces to
\begin{mathletters}
\begin{eqnarray}
{\dot T_{0}^{0}}
+ T_{0}^{0} \Bigl(\frac{\dot a}{a} + \frac{\dot b}{b} +
\frac{\dot c}{c}\Bigr) + T_{1}^{1}\frac{\dot a}{a}
+T_{2}^{2}\frac{\dot b}{b} + T_{3}^{3}\frac{\dot c}{c} &=& 0, \label{cons}\\ 
8 \pi {\dot G} T_{0}^{0}  - {\dot \Lambda} &=& 0, \label{lambda}
\end{eqnarray}
\end{mathletters}

Let us now go back to the spinor field equations (\ref{speq}). 
Using the equalities~\cite{weinberg,brill}
$$ g_{\mu \nu} (x)= e_{\mu}^{a}(x) e_{\nu}^{b}(x) \eta_{ab},
\qquad \gamma_\mu(x)\,=\,e_{\mu}^{a}(x)\bar\gamma^a,$$ 
where $\eta_{ab}= {\rm diag}(1,-1,-1,-1)$,
$\bg_\alpha$ are the Dirac matrices of Minkowski space and
$e_{\mu}^{a}(x)$ are the set of tetradic four-vectors, we obtain 
the Dirac matrices $\gamma^\mu(x)$ of B-I space-time
$$ \gamma^0=\bg^0,\quad \gamma^1 =\bg^1 /a(t),\quad 
\gamma^2= \bg^2 /b(t),\quad \gamma^3 = \bg^3 /c(t), $$
$$ \gamma_0=\bg_0,\quad \gamma_1 =\bg_1 a(t),\quad 
\gamma_2= \bg_2 b(t),\quad \gamma_3 = \bg_3 c(t). $$
The $\Gamma_\mu(x)$ matrices are defined by the equality $$\Gamma_\mu (x)= 
\frac{1}{4}g_{\rho\sigma}(x)\bigl(\pr_\mu e_{\delta}^{b}e_{b}^{\rho} 
- \Gamma_{\mu\delta}^{\rho}\bigr)\gamma^\sigma\gamma^\delta, $$ 
which gives
\begin{equation} 
\Gamma_0=0, \quad 
\Gamma_1=\frac{1}{2}\dot a(t) \bg^1 \bg^0, \quad 
\Gamma_2=\frac{1}{2}\dot b(t) \bg^2 \bg^0, \quad 
\Gamma_3=\frac{1}{2}\dot c(t) \bg^3 \bg^0.
\label{afc}
\end{equation}
Flat space-time matrices we choose in the form~\cite{bogoliubov}
\begin{eqnarray}
\bg^0&=&\left(\begin{array}{cccc}1&0&0&0\\0&1&0&0\\
0&0&-1&0\\0&0&0&-1\end{array}\right), \quad
\bg^1\,=\,\left(\begin{array}{cccc}0&0&0&1\\0&0&1&0\\
0&-1&0&0\\-1&0&0&0\end{array}\right), \nonumber\\
\bg^2&=&\left(\begin{array}{cccc}0&0&0&-i\\0&0&i&0\\
0&i&0&0\\-i&0&0&0\end{array}\right), \quad
\bg^3\,=\,\left(\begin{array}{cccc}0&0&1&0\\0&0&0&-1\\
-1&0&0&0\\0&1&0&0\end{array}\right).  \nonumber
\end{eqnarray}
Defining $\gamma^5$ as follows,
\begin{eqnarray}
\gamma^5&=&-\frac{i}{4} E_{\mu\nu\sigma\rho}\gamma^\mu\gamma^\nu
\gamma^\sigma\gamma^\rho, \quad E_{\mu\nu\sigma\rho}= \sqrt{-g}
\ve_{\mu\nu\sigma\rho}, \quad \ve_{0123}=1,\nonumber \\
\gamma^5&=&-i\sqrt{-g} \gamma^0 \gamma^1 \gamma^2 \gamma^3 
\,=\,-i\bg^0\bg^1\bg^2\bg^3 =
\bg^5, \nonumber
\end{eqnarray}
we obtain
\begin{eqnarray}
\bg^5&=&\left(\begin{array}{cccc}0&0&-1&0\\0&0&0&-1\\
-1&0&0&0\\0&-1&0&0\end{array}\right).\nonumber
\end{eqnarray}
Defining 
\begin{equation}
\label{tau}
\tau(t)=a(t)b(t)c(t) = \sqrt{-g},
\end{equation}
we rewrite the equation (~\ref{speq1}) together  with (\ref{nab}) 
and (\ref{afc}) 
\begin{equation} 
i\bg^0 
\biggl(\frac{\pr}{\pr t} +\frac{{\dot \tau}}{2 \tau} \biggr) \psi 
- \Phi \psi + i {\cG} \gamma^5 \psi = 0. \label{speq11}
\end{equation}  
Defining $U_j(v)= \sqrt{\tau}\,\psi_j(t)$, where 
$j=1,2,3,4,$ and $v = \int\,{\cG} dt$ from 
(\ref{speq11}) one deduces the following system of equations  
\begin{mathletters}
\label{U}
\begin{eqnarray} 
\left. \begin{array}{c}
U_{1}^{\prime} + i (\Phi/{\cG}) U_{1} - U_{3} = 0, \\
U_{3}^{\prime} - i (\Phi/{\cG}) U_{3} + U_{1} = 0, 
\end{array}\right\}\\
\left. \begin{array}{c}
U_{2}^{\prime} + i (\Phi/{\cG}) U_{2} - U_{4} = 0, \\
U_{4}^{\prime} - i (\Phi/{\cG}) U_{4} + U_{2} = 0, 
\end{array}\right\}
\end{eqnarray} 
\end{mathletters}
where  prime denotes differentiation with respect to $v$.

From (\ref{speq}) one can write the equations for 
$ S = \bp \psi, \quad P = i \bp \gamma^5 \psi$ and    
$A = \bp \bg^5 \bg^0 \psi$
\begin{mathletters}
\label{inv}
\begin{eqnarray}
{\dot S_0} - 2 {\cG}\, A_0 &=& 0, \label{S0}\\
{\dot P_0} - 2 \Phi\, A_0 &=& 0, \label{P0}\\
{\dot A_0} + 2 \Phi\, P_0 + 2 {\cG} S_0 &=& 0, \label{A0} 
\end{eqnarray}
\end{mathletters}
where $S_0 = \tau S, \quad P_0 = \tau P$, and $ A_0 = \tau A$,
leading to the following relation
\begin{equation}
S^2 + P^2 + A^2 =  C^2/ \tau^2, \qquad C^2 = \mbox{const.}
\end{equation}
 
Let us now solve the Einstein's equations. To do it we first write the 
expressions for the components of the energy-momentum tensor explicitly. 
Using the property of flat space-time Dirac matrices and the explicit 
form of covariant derivative $\nabla_\mu$, one can easily find
\begin{equation}
T_{0}^{0}= m\,S - \lambda F(I,\,J) + \ve, \quad 
T_{1}^{1}=T_{2}^{2}=T_{3}^{3}= {\cD} S\,+\, {\cG} P 
- \lambda F(I,\,J) -\,p. 
\label{temc}
\end{equation}
Summation of Einstein equations (\ref{11}), (\ref{22}),(\ref{33}) and 
(\ref{00}) multiplied by 3 gives
\begin{equation}
\frac{\ddot 
\tau}{\tau}= 12 \pi G \Bigl(T_{1}^{1}+T_{0}^{0}\Bigr) - 3 \Lambda =
12 \pi G\,\bigl(m S + {\cD} S + {\cG} P - 2 \lambda F(I,\,J) \,+\ve - 
\,p\bigr) - 3 \Lambda.
\label{dtau}
\end{equation} 

Let us express $a, b, c$ through $\tau$. For this we notice that
subtraction of Einstein equations  (\ref{22}) and (\ref{11})  leads  to  
the equation 
\begin{equation}
\frac{\ddot a}{a}-\frac{\ddot b}{b}+\frac{\dot a \dot c}{ac}- 
\frac{\dot b \dot c}{bc}= \frac{d}{dt}\biggl(\frac{\dot a}{a}- 
\frac{\dot b}{b}\biggr)+\biggl(\frac{\dot a}{a}- \frac{\dot b}{b} \biggr) 
\biggl (\frac{\dot a}{a}+\frac{\dot b}{b}+ \frac{\dot c}{c}\biggr)= 0. 
\end{equation} 
with the solution
\begin{equation}
\frac{a}{b}= D_1 \mbox{exp} \biggl(X_1 \int \frac{dt}{\tau}\biggr), \quad 
D_1=\mbox{const.}, \quad X_1= \mbox{const.} 
\label{ab}
\end{equation}
Analogically, one finds
\begin{equation} 
\frac{a}{c}= D_2 \mbox{exp} \biggl(X_2 \int \frac{dt}{\tau}\biggr), \quad 
\frac{b}{c}= D_3 \mbox{exp} \biggl(X_3 \int \frac{dt}{\tau}\biggr),  
\label{ac}
\end{equation}
where $D_2, D_3, X_2, X_3 $ are integration constants. In view of
(\ref{tau}) we find the following functional dependence between the 
constants $D_1,\, D_2,\, D_3,\, X_1,\, X_2,\, X_3 $:  
$$ D_2=D_1\, D_3, \qquad X_2= X_1\,+\,X_3.$$
Finally, from (\ref{ab}) and (\ref{ac}) we write $a(t), b(t)$, and $c(t)$ 
in the explicit form  
\begin{mathletters}
\label{abc}
\begin{eqnarray} 
a(t) &=& 
(D_{1}^{2}D_{3})^{1/3}\tau^{1/3}\mbox{exp}\biggl[\frac{2X_1 
+X_3}{3} \int\,\frac{dt}{\tau (t)} \biggr], \label{a} \\
b(t) &=& 
(D_{1}^{-1}D_{3})^{1/3}\tau^{1/3}\mbox{exp}\biggl[-\frac{X_1 
-X_3}{3} \int\,\frac{dt}{\tau (t)} \biggr], \label{b}\\
c(t) &=& 
(D_{1}D_{3}^{2})^{-1/3}\tau^{1/3}\mbox{exp}\biggl[-\frac{X_1 
+2X_3}{3} \int\,\frac{dt}{\tau (t)} \biggr]. \label{c}
\end{eqnarray}
\end{mathletters}
Thus the system of Einstein's equations is completely integrated. Let us now
go back to the equations (\ref{cons}) and (\ref{lambda}). In view of 
(\ref{temc}) and (\ref{tau}) we find from (\ref{cons}) we find
\begin{equation}
{\dot \ve} + (\ve + p) \frac{\dot \tau}{\tau}
+ \Phi {\dot S}_0 - {\cG} {\dot P}_0 = 0.
\label{cons1}
\end{equation}
On the other hand from (\ref{S0}) and (\ref{P0}) we have 
$$\Phi {\dot S}_0 - {\cG} {\dot P}_0 = 0.$$
Taking this into account and also the equation of state
$p\,=\,\zeta\,\ve, \qquad 0\,\le\, \zeta\,\le\,1$
from (\ref{cons1}) we find 
\begin{equation}
\ve = \frac{\ve_0}{\tau^{1+\zeta}}, \quad
p = \frac{\zeta\,\ve_0}{\tau^{1+\zeta}},
\label{ve}
\end{equation}
where $\ve_0$ is the integration constant.
Let us now define $G$. Taking into account that 
${\dot G} = {\dot \tau} {\pr G}/{\pr \tau}$ and 
${\dot \Lambda} = {\dot \tau} {\pr \Lambda}/{\pr \tau}$
we rewrite (\ref{lambda}) as
\begin{equation}
8 \pi  T_{0}^{0} \frac{\pr G}{\pr \tau} = \frac{\pr \Lambda}{\pr \tau}. 
\label{gdot}
\end{equation}
On the other hand, inserting $a,\,b,\,c$ from (\ref{abc}) into (\ref{00})
we obtain
\begin{equation}
8 \pi T_{0}^{0} G =  \frac{{\dot \tau}^2}{3 \tau^2}
- \frac{{\cal X}}{3 \tau^2} + \Lambda. \label{g}
\end{equation}
where ${\cal X} = X_{1}^{2} + X_1 X_3 + X_{3}^{2}$.
Dividing (\ref{gdot}) by (\ref{g}) we find the following equation for $G$
\begin{equation}
\frac{{\pr G}/{\pr \tau}}{G} = \frac{3 \tau^2 {{\pr \Lambda}/{\pr \tau}}}
{ {\dot \tau}^2 - {\cal X} + 3 \tau^2 \Lambda}.
\label{gdg}
\end{equation}
Now, $\Lambda$ is a given function of $\tau$, namely, 
$\Lambda = \Lambda_0/\tau^2$ as well as $T_{1}^{1}$ and $T_{0}^{0}$.
Then (\ref{dtau}), multiplied by $2 {\dot \tau}$ can be written as
\begin{equation}
2 {\dot \tau}\,{\ddot \tau} = \bigl[2\bigl(12\pi G ( T_{1}^{1} + T_{0}^{0})
- 3 \Lambda \bigr) \tau\bigr] {\dot \tau} = \Psi(\tau) {\dot \tau} = 
{\dot \Psi}(\tau)            
\label{taug}
\end{equation} 
Solution to the equation (\ref{taug}) we write in quadrature
\begin{equation}
\int\,\frac{d \tau}{\sqrt{\Psi (\tau)}} = t.
\label{quad}
\end{equation}
Giving the explicit form of $F(I,\,J)$, from (\ref{quad}) one finds 
concrete function $\tau(t)$. Once the value of $\tau$ is obtained, one can 
get expressions for components $\psi_j(t), \quad j = 1, 2, 3, 4.$

In what follows, we analyze the solutions obtained previously. In 
~\cite{saha1} we gave a detailed analysis of the problem for different
$F(I,\,J)$. Here we give a brief account of that.

Setting $F = F(I)$, i.e. when ${\cG} = 0$ from (\ref{S0}) one finds
\begin{equation}
S = \frac{C_0}{\tau}, \quad C_0= \mbox{const.}
\end{equation}
For the spinor field in this case we obtain
\begin{eqnarray} 
\psi_{r}(t) = (C_r/\sqrt{\tau})\,e^{-i\Omega}, \quad r=1,2, \quad 
\psi_{l}(t) = (C_l/\sqrt{\tau})\,e^{i\Omega}, \quad l=3,4.  
\end{eqnarray} 
where $\Omega = \int \Phi dt$ and 
$ C_r$ and $C_l$ are integration constants such that 
$C_0 = C_{1}^{2}+C_{2}^{2}-C_{3}^{2}-C_{4}^{2}.$

If one sets $F = F(J)$ and puts $m =0$, i.e. when ${\Phi} = 0$ from 
(\ref{P0}) one finds
\begin{equation}
P(t) = \frac{D_0}{\tau}, \quad D_0=\,\mbox{const.}
\end{equation}
In this case for spinor field we obtain
\begin{eqnarray}
\left.\begin{array}{c}
\psi_1 = (1/\sqrt{\tau}) (D_1 e^{i v} + iD_3 e^{-i v}), \quad
\psi_2 = (1/\sqrt{\tau}) (D_2 e^{i v} + iD_4 e^{-i v}), \\
\psi_3 = (1/\sqrt{\tau}) (iD_1 e^{i v} + D_3 e^{-i v}), \quad
\psi_4 = (1/\sqrt{\tau}) (iD_2 e^{i v} + D_4 e^{-i v}).
\end{array}\right\}
\end{eqnarray}                
where $D_0=2\,(D_{1}^{2} + D_{2}^{2} - D_{3}^{2} -D_{4}^{2}).$

Let us note that, in the unified nonlinear 
spinor theory of Heisenberg, the massive term remains absent, and according 
to Heisenberg, the particle mass should be obtained as a result of 
quantization of spinor prematter~\cite{heisenberg}. In the nonlinear 
generalization of classical field equations, the massive term does not 
possess the significance that it possesses in the linear one, as it by 
no means defines total energy (or mass) of the nonlinear field system. 
So our consideration massless spinor field is justified.

Another choice of nonlinear term is
$F = F(K_{\pm}), \quad K_{+} = I + J = I_v = -I_A, \quad
K_{-} = I - J = I_T$.
In the case of massless NLSF one finds
\begin{equation}
S^2 \pm P^2 = \frac{D_{\pm}}{\tau}.
\end{equation}

In all the cases mentioned above we mainly found $\tau = \alpha t$ for
small $t$ guarantying anisotropic behavior of the universe at initial
state, while $\tau = \beta t^2$ as $t \to \infty$ which is in accord
with present day isotropic state. Note that this result was obtained for
$G$ constant. Here $\alpha$ and $\beta$ are constants. As one sees, for  
$\tau = \alpha t$ as $t \to 0$, the solutions of spinor field are
initially singular. But for some special cases, it is possible to
obtain the solutions those are initially regular~\cite{saha1,saha2}, 
but it violates the dominant energy condition in the Hawking-Penrose 
theorem~\cite{zeldovich}. Note that one comes to the analogical conclusion 
choosing $F= F(K), \quad K=IJ.$
 
Now, setting  $\Lambda = \Lambda_0/ \tau^2$ from (\ref{gdg}) we find
$$ G = C/\tau^{6 \Lambda_0/(\alpha^2 - {\cal X} + 3 \Lambda_0)}, \qquad
\tau = \alpha t, \quad C = {\rm const.},$$  
and
$$ G = D \Bigl(\frac{4 \beta \tau}{4 \beta \tau - {\cal X} + 3 \Lambda_0}
\Bigr)^{6 \Lambda_0/({\cal X} - 3 \Lambda_0)}, \qquad
\tau = \beta t^2, \quad D = {\rm const.}.$$

If we consider $\Lambda = \Lambda_0/\tau^2$ and $G$= constant, then
the conservation law $T_{\nu;\mu}^{\mu} = 0$ doesn't hold separately, as 
in that case ~(\ref{lambda}) leads to $\Lambda = {\rm const.}$, which
contradicts our assumption. In this case from (\ref{divI}) we find
\begin{equation}
{\dot \ve} + (1 + \zeta)\frac{\dot \tau}{\tau} = 
-\frac{2 \Lambda_0 {\dot \tau}}{\tau^3}
\label{gc}
\end{equation}
with the solution
\begin{equation}
\ve = \frac{2\Lambda_0}{1 - \zeta} \frac{1}{\tau^2}
\label{ves}
\end{equation}
Setting $F = K^n$ with $K=\{I,\,J,\,(I \pm J),\,IJ\}$ from (\ref{quad}) we
conclude that even in presence of time dependent $\Lambda$ in the Einstein's
equation perfect fluid plays no role at the early stage of expansion
as well as isotropization of BI universe leaving it to the nonlinear 
spinor term in (\ref{splag}) which confirms our claim made 
in~\cite{saha1,saha2}.  

Finally, we see what heppens with the system in absence of spinor field.
As one sees, in this case the relation $\ve = \ve_0/\tau^{1+\zeta}$ takes
place. Given $\Lambda = \Lambda_0/\tau^2$ from (\ref{lambda}) for $G$
one finds
\begin{equation}
G = \frac{\Lambda_0}{4\pi \ve_0 (1 -\zeta)}\,\frac{1}{\tau^{(1-\zeta)}}
\label{Gp}
\end{equation} 

\section{Conclusions}

Exact solutions to the NLSF equations have been obtained for the 
nonlinear terms being arbitrary functions of the invariant 
$I = S^2$ and $J = P^2$, where $S=\bar \psi \psi$ and 
$P= i \bar \psi \gamma^5 \psi$ are the real bilinear forms of spinor 
field, for B-I space-time. It has been shown that introduction of
time dependent $\Lambda$ term in Einstein's equation and consideration
of gravitational constant to be a function of time do not effect the 
initial singularity and asymptotic isotropization process which is dominated
by the nonlinear spinor term in the Lagrangian. It has also been
shown that the results remain unchanged even in the case when the B-I 
space-time is filled with perfect fluid.

\end{document}